# Correlation of $T_c$ with Crystal Chemical Parameters in High-$T_c$ Cuprates, Diborides and Borocarbides: Concept of Arrangement and Function of Layered Superconductors

L.M. Volkova[1], S.A. Polyshchuk[1], F.E. Herbeck[2]

[1] Institute of Chemistry, Far Eastern Branch of Russian Academy of Science, 690022 Vladivostok, Russia

[2] Institute of Automation and Control Processes, Far Eastern Branch of Russian Academy of Science, 690041 Vladivostok, Russia.

**Abstract**

The problems of participation of crystal chemical factors in creating of conditions to occurrence of a superconducting state in layered superconductors are considered. On the basis of the experimental data, given in the literature, new correlations of $T_c$ with relation of crystal chemical parameters in high-$T_c$ cuprates, diborides and borocarbides are established, the critical crystal chemical parameters controlling $T_c$ are selected, and their functions in creating of prerequisites to the occurrence of superconductivity are determined. The similar character of the correlations allows to make conclusion about the common arrangement and function of layered superconductors irrespective of a nature of their superconductivity. The distinctions of crystal chemical characteristic of low-temperature superconductors from high-temperature ones are considered as the factors, which suppress superconductivity. It was shown, that the superconducting process is concentrated in an anisotropic three-dimensional fragment - sandwich, which arrangement and function allows to have stream of charge carriers and besides the space for its carry and resource of effect on this stream. If in a superconductor there are several such fragments, the $T_c$ value of compound is defined by intrinsic $T_c$ of that fragment, whose parameters are closer to optimal. The crystal chemical concept of arrangement and function of layered superconducting materials is advanced. It shows what conditions are necessary in addition to ones defining metal-insulator transition for appearance of superconductivity. The perspectives of the extension of this concept on superconductors with one-dimensional and three-dimensional structures and possibility of its application for prediction of new superconductors are discussed.

# Contents

# 1. Introduction

One of the important problems of superconductivity is the definition of a role of material crystal chemical characteristics in realization of transition in a superconducting state and in rise of temperature of this transition. The experimental proofs of correlation of superconducting properties with crystal chemical parameters of compounds were the stimulus for development of investigation in this direction. High-$T_c$ cuprates are the most obvious case of this correlation.

It is common practice to connect the temperature of transition in a superconducting state, like the majority of properties high-$T_c$ cuprates, predominantly with concentration of charge carriers [1-5] and, accordingly, with lattice parameter *a* or interatomic distances d(Cu-O) in a $CuO_2$ plane [6-9], containing these carriers, and with number of $CuO_2$ planes in a superconductor [8-11]. These positions are the basic both to simulation of new superconductors, and to construction of various crystal chemical dependences. However dependences, obtained on this basis, are not universal and are fulfilled only in limits of several families of high-$T_c$.

For solution of this problem we have installed common features of layered superconducting compounds, and then have selected specialities, which make modify this phenomenon. Experimental data and our researches show that the concentration of charge carriers is not the unique parameter, which defines the transition of substance in a superconducting state and value of transition temperature. It was selected the critical crystal chemical parameters from a rather extensive set offered by experimenters, and were found between the relations and functional connections between them.

The main idea of researches consist in that for occurrence of a superconductivity in layered compounds answers not a separate plane containing charge carriers but an anisotropic three-dimensional structural fragment - sandwich, which crystal chemical parameters correlate with $T_c$. Firstly, was confirmed on the basis of correlation in superconductors with one $CuO_2$ plane, located between two planes from ions only one sort - Ba ions [12]. This approach has allowed to state the correlations in other classes of high-temperature and low-temperature layered superconductors [13-16], to find the critical crystal chemical parameters controlling temperature of transition, and on this basis to formulate the concept of arrangement and function of layered superconducting materials [17].

In this paper we shall consider similarity of superconducting fragments and crystal chemical correlations of Tc with crystal chemical parameters in high-Tc cuprates, diborides and borocarbides, critical crystal chemical parameters controlling Tc in this superconductors and concept of arrangement and function of layered superconductors. Besides we shall discuss a number of actual problems, such as:

- Is there is a dependence of $T_c$ from number of $CuO_2$ planes?
- Why $T_c$ of $MgB_2$ is the highest in a number of diborides?
- Is there the competition between magnetism and superconductivity in layered nickel borocarbides?

## 2. CORRELATION BETWEEN $T_C$ AND THE CRYSTAL CHEMICAL PARAMETERS IN LAYERED SUPERCONDUCTORS

### 2.1. $A_2CuO_2$ sandwich - key to a superconductivity

The structure fragment common to all high-$T_c$ cuprates - multilayer perovskite-like sandwich $A_{n+1}(CuO_2)_n$ consisting of $CuO_2$ planes between the planes of cations A (A = Ca, Sr, Ba, Y, Ln, etc.). The absence of oxygen atoms in the planes of cations A or their removal from the $CuO_2$ plane to the space beyond the planes of cations A (e.g., by 0.8 Å in Hg-1212 and Hg-1223 and by 0.3 Å in Y-123) as a result of the strong Jahn-Teller distortion of the copper environment leads to location of the $CuO_2$ plane between the positively charged planes; this is the chief distinction of the structure of cuprate high-$T_c$ cuprates from the perovskite-like structures of other elements. With the discovery of infinite-layer high-$T_c$ cuprates, $Ca_{1-x}Sr_xCuO_2$ and $Sr_{n+1}Cu_nO_{2n+1+y}$ [18, 19], it has become evident that the $A_2CuO_2$ sandwich is the simplest structure fragment responsible for the high-$T_c$ superconducting properties. The cation sublattice of this sandwich comprises a square planar Cu layer, with the A cations located above and below the centers of the squares. Critical crystal chemical parameters controlling $T_c$ of high-$T_c$ cuprates should obviously be sought among the parameters of the fragment.

The size effect of cations A on $T_c$ is known since the discovery of of high-$T_c$ cuprates [20, 21]. Moreover, the size of the doping cations (for fixed hole concentration and lattice parameter) also affects $T_c$ [22]. For $(Ln_{1-x}M_x)_2CuO_4$, it was shown that $T_c$ changes as a linear function of the difference between the sizes of cations A, decreasing oppositely with the latter. The sensitivity of superconductivity to the distances from the $CuO_2$ plane to the adjacent planes of cations A [$d(CuO_2$-A)] was reported in [23-30] and $d$(A-O) in [31]. $T_c$ of the $Ln_{1.9}Sr_{0.1}CuO_4$ superconductor was doubled (from 25 to 49 K) by strong epitaxial compression, which decreased the $a$ parameter by 0.02 Å and increased the $c$ parameter by 0.1 Å [32]. Hence it follows that $T_c$ increases oppositely with $d$(Cu-Cu) in the plane and in direct proportion to $d(CuO_2$-A). In addition to the size of cation A and its distance to the $CuO_2$ plane, the factor affecting $T_c$ is the charge of this cation. $T_c$ is higher in those optimally doped cuprates (with the same number of $CuO_2$ planes) where the charge of cations A is smaller. For example, $T_c$ of $YBa_2Cu3O_{7-\delta}$ is 93 K [33], and that of $HgBa_2CaCu2O_{6+\delta}$ having $Ca^{2+}$ instead of $Y^{3+}$ is 128 K [34], Variation of $T_c$ with an electric field created by the charges of cations A was noted in [35].

We have shown [12-17], that for occurrence of a superconductivity it is fundamental importance not only a separate $CuO_2$ plane with charge carriers, but the extended structural fragment - sandwich, which arrangement and function allows to have stream of charge carriers and besides the space for its carry and resource of effect on this stream.

In 2.2 and 2.3 we examine the correlation between the $T_c$ and the parameters of $A_2CuO_2$ sandwich in high-$T_c$ cuprates.

## 2.2. High-$T_c$ cuprates with one of CuO$_2$ plane

The Hg-1201 and Tl-2201 phases were chosen for our analysis because, in their structures, the layers adjacent to the $CuO_2$ planes are, as a rule, made up of Ba atoms only, which excludes the effect on the A cation on the interatomic distances of interest.

We examined the $T_c$'s and structural parameters of 42 Hg-1201 and 13 Tl-2201 samples [36-57]. The room-temperature structures of these samples were determined by powder x-ray (20 samples) and neutron (27 samples) diffraction techniques, single-crystal x-ray diffraction (7 samples), and synchrotron x-ray powder diffraction (1 sample). The most marked changes are observed in the Ba-Ba distance (Δ = 0.145 Å in Hg-1201 and Δ = 0.067 Å in Tl-2201), while the changes in the Cu-Ba distance are insignificant (Δ = 0.03 Å).

In these phases, the layers are stacked in the following sequences:

Hg-1201: ...-$(HgO_\delta)^{(2 - 2\delta)+}(BaO)^0(CuO_2)^{(2 - 2\delta)-}(BaO)^0(HgO_\delta)^{(2 - 2\delta)+}$-...
Tl-2201: ...-$(Tl_2O_{2 + \delta})^{(2 - 2\delta)+}(BaO)^0(CuO_2)^{(2 - 2\delta)-}(BaO)^0(Tl_2O_{2 + \delta})^{(2 - 2\delta)+}$-...

The oxygens in the BaO plane are not level with the Ba atoms but are separated from the $CuO_2$ plane by a distance 0.69-0.91 Å larger. In view of this, we consider the -Ba-$CuO_2$-Ba- structure component, in which the $CuO_2$ plane, containing positively charged carriers, is sandwiched between positively charged layers of Ba cations.

Examining the correlations between $T_c$ and various structural parameters derived in earlier studies for a small number of samples (a total of 55 Hg-1201 and Tl-2201 samples), we revealed that $T_c$ is not straightforwardly correlated with the following parameters: (1) lattice parameter $a$ (or the Cu-Cu distance along the diagonal of the $ab$ plane, or the Cu-$O_p$ bond length; (2) Cu-$O_{ap}$ distance; (3) BVS Cu[58]; (4) Cu-Ba distance (Fig. 1). In numerous studies of $HgBa_2CuO_{4+\delta}$ and $TlBa_2CaCu_2O_7$, the Cu-$O_{ap}$ distance was shown to correlate with the amount of excess oxygen and, hence, the amount of holes in the $CuO_2$ plane. The lack of such a correlation between data obtained in different studies was attributed to difficulties in determining the exact content of excess oxygen and concentration of holes in the $CuO_2$ plane. In view of this, it was suggested to consider the correlation between $T_c$ and the position of the Ba atom ($z$ parameter of Ba in Hg-1201 [42] and Tl-1212 [59, 60]; distance from the Ba atom to the $HgO_\delta$ plane (HgO-Ba) or Hg-Ba distance [40]), which is very sensitive to the oxygen stoichiometry: the HgO-Ba and, accordingly, Hg-Ba distances decrease with increasing oxygen content. We checked this correlation for a large number (42) of Hg-1201 samples and found that $T_c$ does not correlate with the oxygen stoichiometry but does correlate with the HgO-Ba and Hg-Ba distances (Figs. 2a, 2b), varying roughly parabolically.

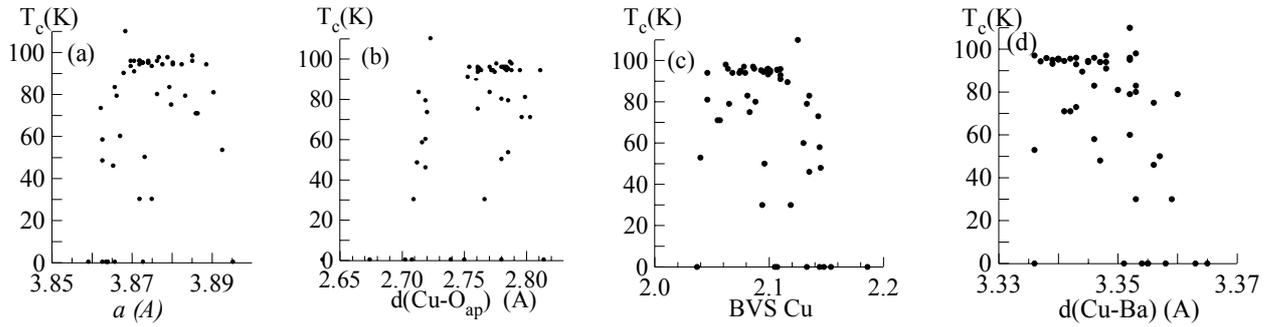

**Fig.1.** $T_c$ as function of $a$ (a), Cu-$O_{ap}$ distance (b), BVS Cu (c), and Cu-Ba distance (d) in Hg-1201 and Tl-2201.

In our opinion, these correlations not only indicate that $T_c$ is connected with the content of excess oxygen, reflected by the position of the Ba atom, but also, and more importantly, provide evidence that the superconducting properties are sensitive to the distance between Ba cations and the $CuO_2$ plane, since the HgO-Ba and Ba-Ba distances are negatively correlated: with increasing $d$(HgO-Ba), $d$(Ba-Ba) decreases (Fig. 3a). The deviations from this relationship are mainly due to the effect of the size factor and the changes in hole concentration associated with nonstoichiometry of the $HgO_\delta$ layer (Hg vacancies; Hg substitutions for Cu, Bi, or Pb; incorporation of $CO_3^{2-}$ groups in the HgO layer; etc.). Combining data for 42 Hg-1201 and 13 Tl-2201 samples, we obtained a better-defined parabolic dependence of $T_c$ on the Ba-Ba distance, equal to twice the separation between the Ba and $CuO_2$ planes (Fig. 2c). Therefore, $T_c$ correlates with the distance from the Ba cation to the cuprate layer. That the superconductivity in Tl-based phases is sensitive to this distance was also pointed out earlier by Simonov and Molchanov [61].

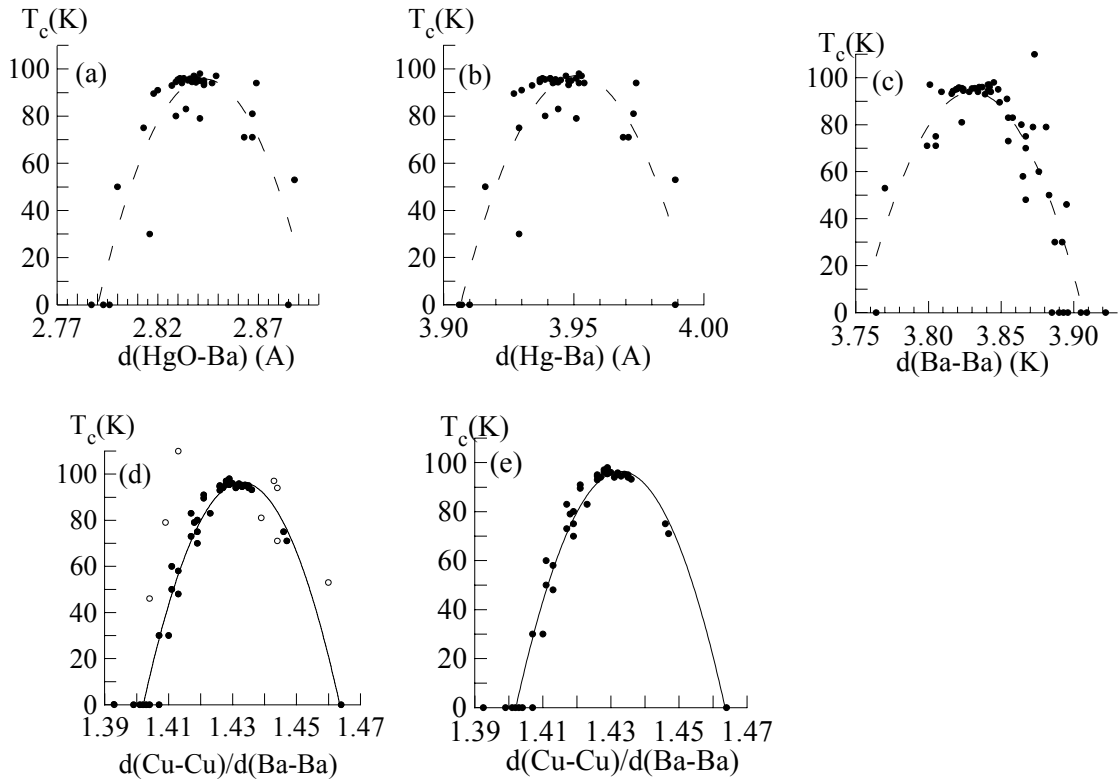

**Fig.2.** $T_c$ as function of (a) HgO-Ba and (b) Hg-Ba distances in Hg-1201 and then (c) Ba-Ba distance and (d, e) the ratio of the Cu-Cu and Ba-Ba distances in Hg-1201 and Tl-2201.

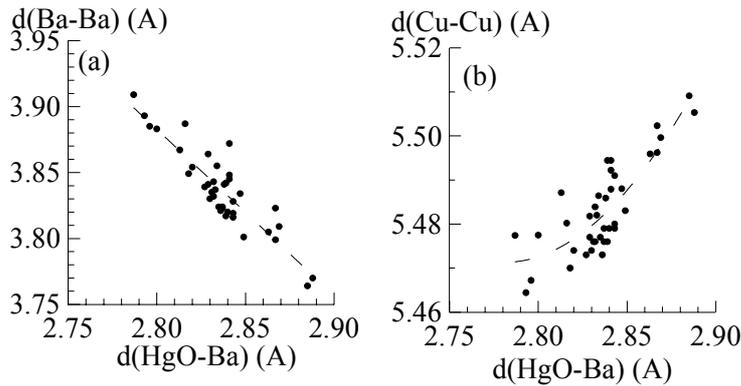

**Fig.3.** HgO-Ba distance as function of the Ba-Ba (a) and Cu-Cu (b) distances in Hg-1201.

At the same time, the deviations of the $T_c$ vs. $d$(Ba-Ba) data from parabolicity suggest that changes in $T_c$ may also be due to both changes in $d$(Cu-Cu) at constant $d$(Ba-Ba) and changes in $d$(Ba-Ba) at constant $d$(Cu-Cu). In addition, there is a tendency for $d$(Cu-Cu) to increase with $d$(HgO-Ba) in Hg-1201 (Fig. 3b). This suggests that $T_c$ depends on the two major parameters of the perovskite sandwiches - the Cu-Cu distance along the diagonal of the $CuO_2$ plane and the separation between the Ba layers. The plot of $T_c$ vs. $J = d$(Cu-Cu)$/d$(Ba-Ba) was found to be closer to parabolicity than the curves considered above (Fig. 2 d). The axis of the parabola thus obtained passes through the value $J = J_0 = 1.431$, which corresponds to the highest $T_c$. Eight out of 55 data points lie far away from the parabola, which can be attributed to experimental errors (Fig. 2d, 2e.).

Thus, our analysis demonstrates that $T_c$ depends on the Cu-Cu and Ba-Ba distances in the cation sublattice of the perovskite layer. All the other atoms only ensure the optimal values of these parameters.

Below, we will consider the relation between the Cu-Cu and Ba-Ba distances and try to reveal factors governing these parameters.

*2.2.1. Factors determining the Cu-Cu and Ba-Ba distances in the us of Hg-1201 and Tl-2201*

An inherent feature of all the high-$T_c$ materials is structural disordering, associated with nonstoichiometry, lattice mismatch between layers, and the presence of charge carriers in the $CuO_2$ plane. The entire data set we use in our analysis was obtained by diffraction techniques and, hence, provides average interatomic distances. Extracting quantitative information from average structural parameters involves many difficulties, which we will try to obviate below.

The appearance of holes on $Cu^{2+}$ ions (transition of $Cu^{2+}$ to $Cu^{3+}$) somewhat reduces the Cu-$O_p$ and Cu-$O_{ap}$ distances. Because of the presence of $Cu^{2+}$ and $Cu^{3+}$, there will be two different Cu-O distances in the $CuO_2$ plane: shorter, between $Cu^{3+}$ and oxygens, and longer, between $Cu^{2+}$ and oxygens. EXAFS spectroscopy of high-$T_c$ superconductors does indicate the presence of at least two Cu-$O_p$ and two Cu-$O_{ap}$ distances. The difference between the longer and shorter Cu-$O_p$ distances lies in the range 0.08-0.12 Å, and that between the longer and shorter Cu-$O_{ap}$ distances ranges between 0.08 and 0.14 Å [62-64].

The presence of oxygens from the $CuO_2$ plane in the nearest neighbor environment of Ba is a necessary condition for the stability of the perovskite block. Let two holes appeared on copper atoms situated at the opposite ends of the diagonal of a square in a $CuO_2$ layer. Then, the compression of the square owing to the shifts of Cu atoms because of the contraction of the four Cu-$O_p$ distances will reduce the Ba-O distances. This will, in turn, result in a shift of the two Ba atoms along the $c$ axis away from the square plane for the Ba-O bond lengths to remain unchanged. On the other hand, the increase in the Ba-Ba distance will be promoted by the increase in the Coulomb repulsion between Ba atoms and two Cu atoms, because of the increase in their charge state from 2+ to 3+, since the Ba-Cu distances (3.34-3.37 Å) are not much longer than the sum of the ionic radii of Ba and Cu (3.2 Å).

The Cu-Cu distances in the cuprate plane and, hence, the Ba-Ba distances may depend not only on the presence of holes but also on the size factor (e.g., the size of M cations in other layers). For example, as a result of substitution of the larger sized ion $Hg^{2+}$ (1.02 Å) for the smaller sized ion $Tl^{3+}$ (0.67 Å), the Cu-Cu distances in Tl-2201 are typically shorter and, accordingly, the Ba-Ba distances are longer than those in Hg-1201. Only in heavily doped Hg-1201 do these distances approach those in Tl-2201. Therefore, it is reasonable to assume that the sterically hindered contraction of the Cu-$O_p$ bonds is compensated for by the larger contraction of the Cu-$O_{ap}$ bonds upon the formation of holes. At first glance, this inference is in conflict with experimental data: the Cu-$O_{ap}$ bonds in Hg-1201 (2.74-2.81 Å) are much longer than those in Tl-2201 (2.64-2.72 Å). But this conflict can be reconciled readily by assuming that all the Hg-1201 samples are doped to a lower level than are the Tl-1201 samples. Figure 4 demonstrates that the Ba-Ba distances tend to decrease with increasing Cu-Cu distance.

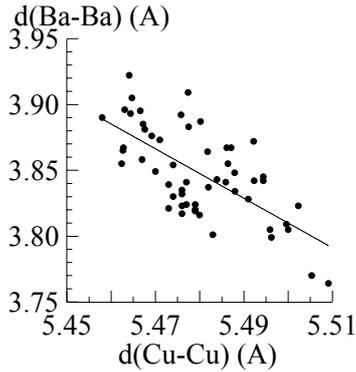

**Fig. 4.** Ba-Ba distance as function of the Cu-Cu distance in Hg-1201 and Tl-2201.

If the crystal lattice is under compression, the decrease in the amount of holes, associated with the unavoidable increase in the Cu-O distance, may be due to the shift of oxygens away from the $CuO_2$ plane (buckling), rather than to an increase in d(Cu-Cu). This will likely change d(Ba-Ba) at constant d(Cu-Cu) and, accordingly, constant *a* and *b*. The buckling of the $CuO_2$ plane was revealed in both bulk samples [65, 66] and heteroepitaxial films [67] subjected to compressive stress.

Thus, we believe that the appearance of holes on the Cu atoms situated at the opposite ends of the diagonal of a square in the cuprate layer must lead to the following structural changes: (1) compression of the square through shifts of the Cu atoms and (2) out-of-plane shifts of the two Ba atoms by the same distance. The shifts of the Ba atoms maintain the equilibrium distance between the Ba atoms and the oxygens of the $CuO_2$ plane and, accordingly, minimize the potential energy; hence, they are independent of the other factors.

Note that the two inequivalent positions of the Ba atoms - one closer to and the other farther away from the $CuO_2$ plane [38] - are averaged over four unit cells sharing the hole, which influences the Ba positions in all of them. For example, if only one Cu cation in the unit cell is in the 3+ state, the Ba atoms will obviously shift toward the $Cu^{2+}$ ions, and the Ba-Ba distance will increase. It seems likely that the average Ba-Ba distance incorporates at least three different Ba-Ba distances: the largest, corresponding to $Cu^{3+}$ ions (holes) situated at the ends of the diagonal of the cell; intermediate (one hole); and the smallest (no holes). A similar picture is expected for *d*(Cu-Cu).

Clearly, in addition to the factors considered, the Ba-Ba and Cu-Cu distances are also influenced by the way in which the holes in the cuprate plane are ordered. Experimental studies of the local structure of Bi-2212, Bi-2201, Tl-2212, Tl-2223, $La_2CuO_{4.1}$, and $La_{1.85}Sr_{0.15}CuO_4$ [62-64] reveal that the $CuO_2$ conducting plane is structurally nonuniform. The holes resulting from doping of the antiferromagnetic dielectric form metallic stripes with dielectric regions in between. Gor'kov and Sokol [68] were the first to assume phase separation. Their ideas were developed further by Emeri *et al.* [69].

The number of short Cu-Cu and, accordingly, long Ba-Ba bonds will depend on the type of ordering. If *N* holes arranged in pairs appear at the ends of diagonals at random, the number of short Cu-Cu and long Ba-Ba distances will be *N/2*. This number will increase to *N* - 1 if the holes will be aligned in a row along the diagonal of the *ab* plane, to *3N/2* - 2 for a double row, to 5N/3 - 3 for a stripe of three rows, etc.

Thus, the local Cu-Cu ($d(Cu^{3+}$-$Cu^{3+})$ and $d(Cu^{2+}$-$Cu^{2+}))$ and Ba-Ba distances in hole-rich and dielectric regions are likely to depend only on geometric factors (size of the M cations and its substituents and the presence of vacancies in the MO layer), hydrostatic pressure, and uniaxial strain. The average Cu-Cu and Ba-Ba distances derived from diffraction measurements are determined by the local distances and also by the structure and concentration of hole-rich regions in the "spin field." For example, if the holes are ordered into stripes, the width of the "spin stripes" and, accordingly, the average Cu-Cu distance will decrease with increasing hole concentration, while the average Ba-Ba distance will increase.

The revealed correlation between $T_c$ and the ratio of the average Cu-Cu and Ba-Ba distances suggests that $T_c$ in Hg-1201 and Tl-2201 depends on the following factors: the distance between holes in the pairs located at the ends of squares in the $CuO_2$ planes, the shape of
the hole-rich regions (stripes or clusters), the distance between these regions, and the distance from the cuprate layer to the A layer, $d$(Ba-Ba)/2. It seems likely that $T_c$ is minimal at short and long distances between hole-rich stripes and rises at intermediate distances.

*2.2.2. Role of the perovskite layer in superconductivity.*

To account for the correlation between $T_c$ and d(Cu-Cu)/d(Ba-Ba), it is important to understand which role the structure of the cation sublattice in the perovskite layer plays in the development of high-$T_c$ superconductivity. At first glance, this structure is not unique: a square (or near-square) planar layer of Cu atoms sandwiched between two planes of Ba cations.

Let us make a few assumptions:

(1) Two holes may appear within the same cell of the cuprate layer only on Cu atoms situated at the ends of a diagonal, i.e., at the maximal possible distance from one another, so that the Coulomb interaction between $Cu^{3+}$ cations would be minimal. Since the two diagonal directions in the cell are equivalent, the hole pair may oscillate around the equilibrium, center position, jumping from one diagonal to the other (rotating through 90°). The diagonal pairing and oscillation of holes are energetically more favourable than other configurations, since they lead to less significant lattice distortions in hole-rich regions.

(2) The formation of hole-rich stripes separated by hole-free regions (Fig. 5), in which $Cu^{2+}$($d^9$, $s = 1/2$) ions order antiferromagnetically, must also be favourable from the viewpoint of symmetry retention and smaller atomic displacements on doping. Clearly, the holes in a stripe can be in one of two equivalent configurations (Fig. 5), residing either at the ends of the diagonals normal to the stripe direction (configuration A) or at the ends of the diagonals parallel to the stripe direction (configuration *B)*. Configuration A, incorporating a double row of holes, will have an excess positive charge, δ+, while configuration *B,* consisting of a single row, will have an excess negative charge, δ-. To maintain electroneutrality, stripes A must alternate with stripes *B*.

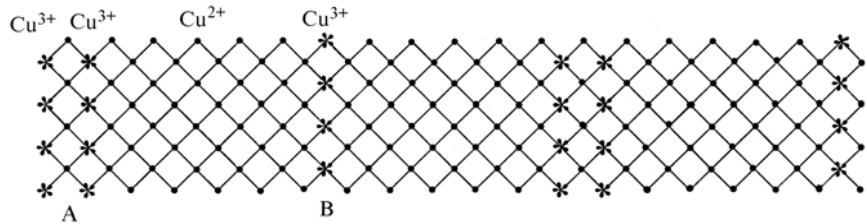

Fig. 5. The scheme of hole arrangement in copper net.

(3) At low temperature, the hindered oscillations of hole pairs can occur only cooperatively, to avoid the presence of holes in neighboring sites for a long time. As a result of the simultaneous rotation of the hole pairs in configuration A through 90°, half of the cells become hole-excess. By contrast, as a result of the rotation of the diagonal hole pairs in configuration *B,* every other cell becomes hole-free. The alternation of oppositely charged stripes and cooperative oscillations of hole pairs will no doubt give rise to hole transfer between neighboring stripes. The rate of hole transfer and the pulse duration, dependent on the average distance between hole-rich regions, are determined by the hole concentration and ordering configuration and are characterized by the d(Cu-Cu)/d(Ba-Ba) ratio.

(4) Along the direction of hole transfer between stripes, spin-flip chains may appear. If the holes then return along the same trajectories, the antiphase ordering will disappear and the original antiferromagnetic ordering will be restored. Spin-flip behavior is suggested by the effect that a moving electron has on the local spin structure [70].

(5) Hole movement stable with respect to various deviations from the trajectory (e.g., localization at apical oxygens) is ensured by the electric field of similarly charged A-cation layers. In the phases under consideration, these are $Ba^{2+}$ layers. An analogous necessity of focusing arises in accelerators of charged particles. The focusing of holes along their trajectory seems to be the major function of the Ba layers. The reduction in the separation between the positively charged Ba layers implies, on the one hand, a decrease in the cross-sectional area of hole-transfer regions and, on the other hand, an increase in the degree of hole focusing.

It seems likely that the holes can be defocused, if there are a few types of A cations differing in size and/or charge, because of the scattering of holes by cations appearing in the transfer region and also because of the increase in the oscillation amplitude.

Hence by correlating $T_c$ with interatomic distances in the structure of Hg-1201 and Tl-2201, we found that the structural parameters which might be responsible for enhancement of $T_c$ are the Cu-Cu distance along the diagonal direction in the $CuO_2$ plane and the Ba-Ba distance between the Ba planes in the cation sublattice of the perovskite layer.

$T_c$ was found to vary parabolically with the ratio of the average Cu-Cu and Ba-Ba distances derived from diffraction measurements. The origin of this correlation is that the cation sublattice of the perovskite layer, containing charge carriers, is responsible for the development of superconductivity.

The average Cu-Cu and Ba-Ba distances in the unstrained structure are determined primarily by the hole concentration and type of ordering and also the dimensions of the M cations in the layers located between the perovskite sandwiches. The effects of hydrostatic pressure and epitaxial strain change these relations.

We suppose that the interatomic distances in quesion reflect the major parameters influencing $T_c$, such as the rate of hole transfer, pulse duration (distance between hole-rich regions), cross-sectional area of hole-transfer regions, degree of focusing of holes along their trajectory, and distance between the holes in an oscillating pair. By optimizing this set of parameters, it will be possible to raise $T_c$.

It seems likely that the attainment of high $T_c$'s depends on the possibility of focusing holes along their trajectory by the electric field of the A-cation layers, which are located closer to the $CuO_2$ plane than are the apical oxygen, because of the strong Jahn-Teller distortion of the copper coordination.

Varying the Cu-Cu and Ba-Ba distances offers the means of enhancing $T_c$.

## 2.3. High-$T_c$ cuprates with n of $CuO_2$ planes

### 2.3.1. *Superconducting transition temperature and number of $CuO_2$ planes*

In main it is accepted to link $T_c$ to one from parameters, for example, another number of holes one or an other interatomic distance or size of cation, and so on. However, the experiments show, that $T_c$ depends on many factors simultaneously, therefore, phases high-$T_c$ cuprates with various cationic composition have as a rule its own dependences of $T_c$ from like parameters. Above, on an example of two simple phases of Hg-1201 and Tl-2201 we have found the solution of this problem. Main, we selected the elementary structural fragment - sandwich $Ba_2(CuO_2)$, which answers for occurrence of a superconductivity, determined the parameters of this fragment effecting on $T_c$, and found a relation between them correlating with $T_c$. We showed, that $T_c$ correlates with the ratio of distances between copper atoms in a plane CuO2 and total of distances from this plane up to two adjacent planes of cations (atoms Ba). We proved that these parameters, except for independent influence on a superconductivity, carry in the fullest information on concentration of charge carriers in a plane $CuO_2$. This correlation is plotted by one curve, and without usage of any normalizing coefficients.

The following purpose is to find the universal dependence of $T_c$ from crystal chemical parameters for all phases of high-$T_c$ cuprates containing some planes $CuO_2$. It was clear, that a key of this search is the ratio between distances d(Cu-Cu) and total of distances between the plane $CuO_2$ and external planes of A cations in $A_2(CuO_2)$ sandwich. However, in these phases contain there are some such sandwiches. There is a question: how the $T_c$ of a superconductor depends on amount of these fragments? Besides, it was necessary to find the characteristic of planes of A - cations, as in phases viewed they have various cationic composition.

For this we analyzed the structural data of 221 compounds with 2-4 $CuO_2$ planes (n = 2 — 4) from following phases: Y - 123, 124, and 247; Hg - 1212, 1213, 1223, 1234, and 2212; Tl - 1212, 1223, 2212, and 2223; Cu - and Cu,C - 1234;$Tl_{0.5}Pb_{0.5}$ - 1212, 1223; $Hg_{0.5}Pb_{0.5}$ - 1212, and so on. The $T_c$ was used (in the main on the onset of transition) and the structural data obtained at room temperature by x-ray and neutron diffraction on pattern (XP and NP) and on single crystals (XSC and NSC). The following estimation showed that the great numbers of data deviated from found relation because in most cases it is difficult to define correctly the cation content doping the A cation planes from pattern experiment. So from 73 doping compounds, whose structure was defined on single crystals, the deviation was observed for 19 compounds: 7 structures were defined in the first 3 year of high-$T_c$ cuprates study, in 11 compound the Al content was proposed [71]. The deviation numbers for the structures fixed by NP method is 40% and XP - 67%. For reference, from 55 compounds of Hg-1201 and Tl-2201 phases with only one kind of atom in A cation planes, the deviation took place in only 9 cases. Hence, 131 compounds with n = 2 — 4 were selected for the plotting of curves (information on structural parameters and $T_c$ of these compounds may be requested from the authors).

The perovskite layer of the phases with several $CuO_2$-planes is divided on n elementary $A_2CuO_2$ fragments according to numbers of $CuO_2$-planes. These fragments even in only one compound by equal $d$(Cu-Cu) distances may differ by $d(CuO_2$-A) ones and by elemental composition of A-cation planes. For plotting of common for all phases dependence of $T_c$ from main crystal chemical parameters of cation sublattice $A_2Cu$ of this fragments we have chosen the characteristics, such as:

1. The $d$(Cu-Cu) distances between Cu atoms along diagonal direction of $CuO_2$-plane ($d(Cu-Cu) = \sqrt{a^2 + b^2}$), because in some high-$T_c$-phases the $a$ and $b$ lattice constant is not equal;

2. "Effective" distances $D_1$ and $D_2$ from $CuO_2$-plane to surface of two adjacent planes of A-cations:

$$D = S[d(CuO_2 - A) - R_A(Z_A/2)] \qquad (1)$$

where d($CuO_2$-A) is the distance from $CuO_2$-plane to plane of A-cations, $R_A$ is radius [72] of A-cation, which content is maximum, $Z_A/2$ – undimensional coefficient to take into account of the electric field of the A-cation charge (it is the ratio of charge A-cation to charge of Ca cation), $S$ is deviation coefficient of parameters of doping cations from parameters of A-cation that forms the plane:

$$S \geq 1, \ S = \overline{R(Z/2)}/R_A(Z_A/2) \text{ or } S = R_A(Z_A/2)/\overline{R(Z/2)} \qquad (2)$$

Here $\overline{R(Z/2)}$ is generalized value, characterized the plane of A-cations:

$$\overline{R(Z/2)} = m_1 R_{A_1}(Z_{A_1}/2) + ... m_n R_{A_n}(Z_{A_n}/2) \qquad (3)$$

where $m_n$ is content of $A_n$-cation in plane, $R_{A_n}$ is radius, $Z_{A_n}/2$ – undimensional coefficient to take into account of the electric field of the $A_n$-cation charge.

From here on we shall use the dependence of $T_c$ from the ratio ($J$) of distance $d$(Cu-Cu) to sum "effective" distances $D_1$ and $D_2$:

$$J = d(Cu-Cu)/(D_1 + D_2) \qquad (4)$$

By investigation of high-$T_c$ cuprates with by some $CuO_2$-planes (n>1) the dependence $T_c(J)$ was plotted at first for the compounds where there is only one type of $A_2CuO_2$ fragments (Fig. 6a). As rule it is the compounds with n=2. Then we considered the compounds with various $A_2CuO_2$ fragments (for example, formed by external (type-I) and internal (type-II) $CuO_2$-planes), that differ by J and $T_c$ values (Fig. 6b). It was found that $T_c$ of a compound is intrinsic $T_c$ of one from its, just that which have higher $T_c$ (Fig. 6c). From such dependence $T_c$(J) it turns out that the high $T_c$ in the superconductors with tree $CuO_2$-planes determines not by number of $CuO_2$-planes but proximity of parameters of one from of $A_2CuO_2$ fragments to optimum. It is important that in the most compounds examined with various types of $A_2CuO_2$ fragments (in 28 from 38) the fragment formed by external $CuO_2$-planes is responsible for $T_c$.

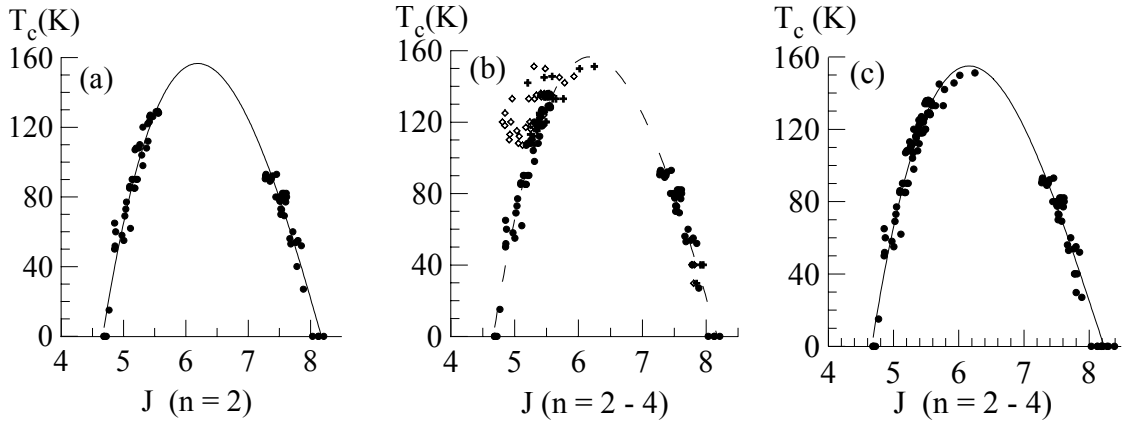

**Fig. 6.** $T_c$ as function of $J$ in high-$T_c$ cuprates (a) with only one type (external) of fragments $A_2CuO_2$, (b) cuprates with two types fragments (external (+) and internal (◊)) were added to cuprates with one type of fragments, (c) fragments with higher $T_c$ were selected from cuprates with two types fragments.

It is interesting to examine the influence of pressure on Hg-1223 phase [73,74]: by low pressure to 0.59 GPa ($T_c$ - 135-136 K, $J_I$ = 5.49 - 5.56, $J_{II}$ =5.40 - 5.46), $T_c$ of compound is, by $T_c(J)$, intrinsic $T_c$ of the type-$I$ fragment, by pressure 4-6 Gpa ($T_c$ = 142-145 K, $J_I$, = 5.20 - 5.59, $J_{II}$ = 5.78 -5.93)—type-II fragment; and by high pressure 8.5 - 9.2 GPa ($T_c$ = 149.9-151.1 K, $J_I$ = 6.02 - 6.25, $J_{II}$ -5.30 — 5.48)—again, intrinsic $T_c$ of type-I fragment. Notice that for $Hg_{0.8}Tl_{0.2}Ba_2Ca_2Cu_3O_{8.33}$ $T_{c(onset)}$ 145 K [75] corresponds to $J_{II}$ (5.79), as $J_I$ (5.46) corresponds to $T_c$ below 130 K; Whereas for explanation of the behavior $T_c$ of Hg-1223 and Hg-1234 under pressure, the authors [76,77] consider that it is necessary to examine as nonequivalent only $CuO_2$ planes but not $A_2CuO_2$ fragments.

Plotting dependence $T_c(J)$ for Hg-1201 and Tl-2201 phases (Fig. 7a), we find that all under investigation high-$T_c$ cuprates are divided only on two groups with its intrinsic dependence $T_c(J)$ (Fig. 7b): the phases formed by single $CuO_2$ plane (n = 1) and by some $CuO_2$ planes $(n > 1)$.

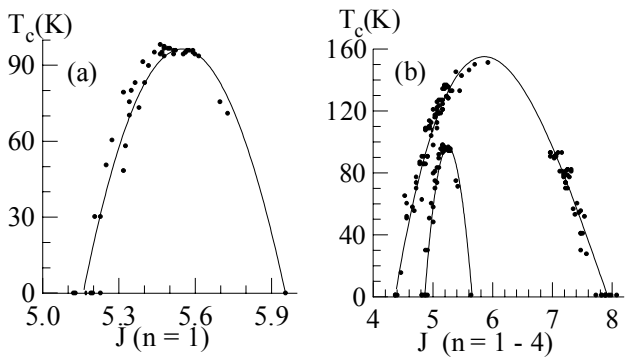

**Fig. 7.** $T_c$ as a function of $J$ in high-$T_c$ cuprates formed (a) by single $CuO_2$ plane (Hg-1201 and Tl-2201 phases) and (b) by 1-4 $CuO_2$ planes.

Graphic plot $T_c(J)$ approximates to parabola. However, the closer approximation of that dependence gives the equation of polynomial of third degree [94% for high-$T_c$ cuprates by
n = 1: $T_c = 63.0222J^3 - 1660J^2 + 12600.3J - 29477.1$ (Fig. 7a) and

97% - by n > 1: $T_c = 9.33369J^3 - 227.846J^2 + 1744.12J - 4124.78$ (Fig. 6c)].
By that relation, the maximal $T_c$ that may be attained in high-$T_c$ cuprates (n > 1) is only 155 K ($J$ = 6,15). For dependence (n > 1) built only on monocrystal data *($T_c = 12.1113J^3 - 286.406J^2 + 2149.29J - 5041$)*, $T_{c,max}$ ($J$ = 6.15) is 162 K, which is only 2° less than the record $T_c$ of 164 K [78], fixed on Hg 1223 phase by pressure.

By this means $T_c$ of high-$T_c$ cuprates is function of the ratio ($J$) of distances between Cu atoms along diagonal direction of $CuO_2$ plane to sum "effective" distances ($D_1 + D_2$) from $CuO_2$ plane to of two adjacent planes of A cations in $A_{n+1}(CuO_2)_n$ layer, taking into account the charge and the size of A cations and doping atoms ($J = d(Cu-Cu)/(D_1+D_2)$). The advantage of that dependence from more early ones is that all phases of high-$T_c$ cuprates are divided into only two groups with its intrinsic dependence $T_c(J)$: the phases formed by one $CuO_2$ plane (n = 1) and by some $CuO_2$ planes (n > 1). The simpler structural fragment where the superconductivity appears probably is not $CuO_2$ plane but anisotropic three-dimensional fragment—sandwich $A_2CuO_2$. $T_c$ of compound with various types of elementary $A_2CuO_2$ fragments is intrinsic $T_c$ of one from its, just those which have higher $T_c$. ~ 164 K is the maximum $T_c$ for high-$T_c$ cuprates on the ground of $CuO_2$ plane.

*2.3.2. Critical crystal chemical parameters controlling $T_c$*

The dependence $T_c$ on charge carriers concentration is expressed in correlation $T_c(J)$ by the ratio of distances between Cu atoms in $CuO_2$ plane and distances between $CuO_2$ plane and A cation planes (see 2.2.1). Moreover, it is shown that these distances have their own effect on $T_c$.

Critical parameters controlling $T_c$, apart from the hole concentration in $CuO_2$-plane, are the distances between Cu-atoms along diagonal direction of $CuO_2$-plane and the parameters characterised the space between of $CuO_2$-plane and A-cation planes in $A_2CuO_2$ sandwich, such as an interval between the surface of the planes, the buckling surface of A cation planes, and the electric fields induced by the A cations and doping cations charges.

The role of interplane space of sandwich $A_2CuO_2$ in the occurrence of superconductivity and of $T_c$ change is very high. There is the dependence of $T_c$ from the distances between Ba atom planes (d(Ba-Ba)) in compounds from Hg-1201 and Tl-2201 (Fig. 2c) (approximation to polynomial 91%). It appeared that in external $A_2CuO_2$ fragments (type-I) the dependence of $T_c$ on the distances to the outer ($D_1$) plane of cations A is unavailable (Fig. 8a), except that at a nearly parabolic dependence between $T_c$ and the "effective" distances from the $CuO_2$ plane to the inner ($D_2$) plane of cations A containing no oxygen atoms (Fig. 8b) were found. The absence of a $T_c(D_1)$ relationship is possibly explained by a decreased effective positive charge on the outer plane of cations A due to the close position of the $O_{ap}$ atoms. The A cation planes in the sandwich seem to be holes focusing on path of motion. The relationship between $T_c$ and the sum of $D_1$ and $D_2$ is even more parabolic (for fragment type-I from 121 compounds and fragment type-II from 10 compounds) (Fig. 8c) (approximation to polynomial for $T_c(D_2)$ is 81% and for $T_c(D_1+D_2)$ is 95%).

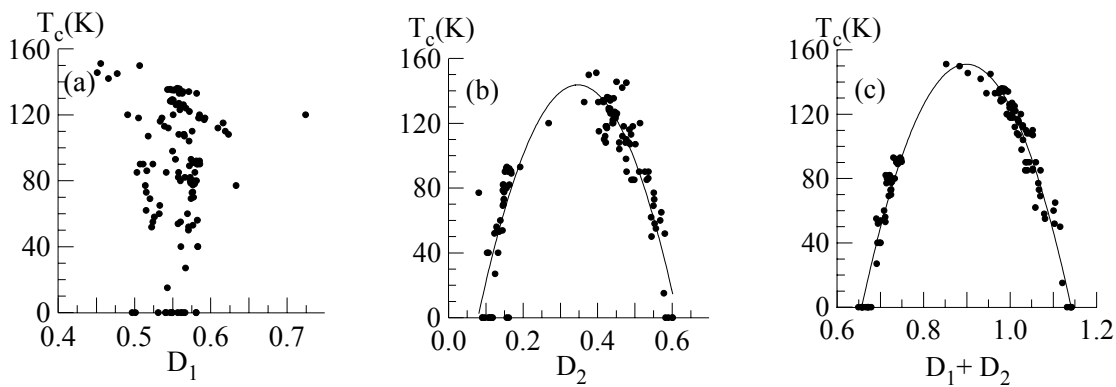

**Fig. 8.** $T_c$ as a function (a) $D_1$, (b) $D_2$ and (c) $D_1+D_2$ in high-$T_c$ cuprates formed by some $CuO_2$ planes

## 4. Conclusion

In this paper the analysis of character of variations of crystal chemical parameters and composition compounds accompanied with transitions in superconducting condition is carried out. The correlations between $T_c$ and ratio of crystal chemical parameters for layered superconductors: high-$T_c$ cuprates, diborides and borocarbides of nickel were established. They have similar character in all three classes of compounds, despite of distinction of a nature of their superconductivity. On the basis of these correlations the critical crystal chemical parameters control of $T_c$, and function of these parameters in occurrence of superconductivity are determined. It is found out, that responsible for occurrence of superconductivity in layered superconductors is anisotropic three-dimensional fragment – sandwich ($A_2(Cu)$ in high-$T_c$ cuprates, $A_2(B_2)$ in diborides and RB(Ni) in borocarbides), which inner plane creates a flow of charge carriers, and the external planes form this flow. Is proved, that at presence in a superconductor of several such fragments, as, for example, in HTSC cuprates, with several $CuO_2$ planes, the $T_c$ of compounds is determined by own value of $T_c$ of that fragment, whose parameters are closer to optimum. The absence a competition between superconductivity and magnetism in nickel borocarbides proved. It is shown, that the problem rising $T_c$ diborides is a problem of nonstoichiometry in a plane of boron in diborides of heavy metals being compounds of variable composition. It is developed the crystal chemical concept of arrangement and functional layered superconducting materials that defines the additional factors for appearance of superconductivity, in additional to conductions determining transition in metal state. It defines the principles of crystal chemical modelling of new superconductors. It is supposed, that this concept can be extended and is applicable to materials with one-and three-dimensional structures.

**ACKNOWLEDGMENTS**
This work was supported by the Russian Foundation for Basic research under grant 00-03-32486 and grant of Far East. Br. Russ. Ac. Sci. The authors thank E. M. Kopnin, R.F. Klevsova, C.B. Borisov, B.B. Bakakin for helpful discussion, and S. A. Magarill and A. N. Sobolev for help in the search for literary experimental data.